\documentstyle[epsf]{mn}

\newif\ifAMStwofonts

\ifoldfss
  \ifCUPmtlplainloaded \else
    \NewTextAlphabet{textbfit} {cmbxti10} {}
    \NewTextAlphabet{textbfss} {cmssbx10} {}
    \NewMathAlphabet{mathbfit} {cmbxti10} {} 
    \NewMathAlphabet{mathbfss} {cmssbx10} {} 
  \fi
  \ifAMStwofonts
    \ifCUPmtlplainloaded \else
      \NewSymbolFont{upmath} {eurm10}
      \NewSymbolFont{AMSa} {msam10}
      \NewMathSymbol{\upi}     {0}{upmath}{19}
      \NewMathSymbol{\umu}     {0}{upmath}{16}
      \NewMathSymbol{\upartial}{0}{upmath}{40}
      \NewMathSymbol{\leqslant}{3}{AMSa}{36}
      \NewMathSymbol{\geqslant}{3}{AMSa}{3E}

      \let\leq=\leqslant \let\le=\leqslant
      \let\geq=\geqslant \let\ge=\geqslant
    \fi
  \fi
\fi 

\ifnfssone
  \newmathalphabet{\mathit}
  \addtoversion{normal}{\mathit}{cmr}{m}{it}
  \addtoversion{bold}{\mathit}{cmr}{bx}{it}
  \newmathalphabet{\mathbfit} 
  \addtoversion{normal}{\mathbfit}{cmr}{bx}{it}
  \addtoversion{bold}{\mathbfit}{cmr}{bx}{it}
  \newmathalphabet{\mathbfss} 
  \addtoversion{normal}{\mathbfss}{cmss}{bx}{n}
  \addtoversion{bold}{\mathbfss}{cmss}{bx}{n}
  \ifAMStwofonts
    \ifCUPmtlplainloaded \else
      \UseAMStwoboldmath
      \makeatletter
      \new@mathgroup\upmath@group
      \define@mathgroup\mv@normal\upmath@group{eur}{m}{n}
      \define@mathgroup\mv@bold\upmath@group{eur}{b}{n}
      \edef\UPM{\hexnumber\upmath@group}
      \new@mathgroup\amsa@group
      \define@mathgroup\mv@normal\amsa@group{msa}{m}{n}
      \define@mathgroup\mv@bold\amsa@group{msa}{m}{n}
      \edef\AMSa{\hexnumber\amsa@group}
      \makeatother
      \mathchardef\upi="0\UPM19
      \mathchardef\umu="0\UPM16
      \mathchardef\upartial="0\UPM40
      \mathchardef\leqslant="3\AMSa36
      \mathchardef\geqslant="3\AMSa3E

      \let\leq=\leqslant \let\le=\leqslant
      \let\geq=\geqslant \let\ge=\geqslant
    \fi
  \fi
\fi 

\ifnfsstwo
  \DeclareMathAlphabet{\mathbfit}{OT1}{cmr}{bx}{it}
  \SetMathAlphabet\mathbfit{bold}{OT1}{cmr}{bx}{it}
  \DeclareMathAlphabet{\mathbfss}{OT1}{cmss}{bx}{n}
  \SetMathAlphabet\mathbfss{bold}{OT1}{cmss}{bx}{n}
  \ifAMStwofonts
    \ifCUPmtlplainloaded \else
      \DeclareSymbolFont{UPM}{U}{eur}{m}{n}
      \SetSymbolFont{UPM}{bold}{U}{eur}{b}{n}
      \DeclareSymbolFont{AMSa}{U}{msa}{m}{n}
      \DeclareMathSymbol{\upi}{0}{UPM}{"19}
      \DeclareMathSymbol{\umu}{0}{UPM}{"16}
      \DeclareMathSymbol{\upartial}{0}{UPM}{"40}
      \DeclareMathSymbol{\leqslant}{3}{AMSa}{"36}
      \DeclareMathSymbol{\geqslant}{3}{AMSa}{"3E}

      \let\leq=\leqslant \let\le=\leqslant
      \let\geq=\geqslant \let\ge=\geqslant
    \fi
  \fi
\fi 

\ifCUPmtlplainloaded \else
  \ifAMStwofonts \else 
    \def\upi{\pi}
    \def\umu{\mu}
    \def\upartial{\partial}
  \fi
\fi

\title[Substructure recovery by 3D Discrete Wavelet Transforms] {Substructure 
recovery by 3D Discrete Wavelet Transforms} 
\author[A. Pagliaro et al.]  
{A. Pagliaro $^{1,2}$, V. Antonuccio-Delogu$^3$\thanks{also:
TAC, Copenhagen, DENMARK}, 
U. Becciani $^3$ and ~ M. Gambera $^2$ \\
$^1$ SRON, Sorbonnelaan, 2 - NL 3584 CA Utrecht, The Netherlands \\
$^2$ Istituto di Astronomia dell'Universit\`a di Catania, 
Citt\'{a} Universitaria, Via S. Sofia 78 - I 95125 Catania, Italy \\
$^3$ Osservatorio Astrofisico di Catania and CNR-GNA, 
Citt\'{a} Universitaria, Via S. Sofia 78 - I 95125 Catania, Italy \\}
\date{}

\begin{document}

\maketitle

\begin{abstract}
We present and discuss a method to identify substructures in combined
angular-redshift samples of galaxies within Clusters. The method relies 
on the use of Discrete Wavelet Transform (hereafter DWT) and has
already been applied to the analysis of the Coma cluster (Gambera et al.
1997). The main new ingredient of our method with respect to previous
studies lies in the fact that we make use of a 3D data set
rather than a 2D. We test
the method on
mock cluster catalogs with spatially localized substructures and on a N-body
simulation. Our main conclusion is that our method is able to
identify the existing substructures provided that: a) the subclumps are detached in 
part or all of the phase space, b) one has a 
statistically significant number
of redshifts, increasing as the distance 
decreases due to redshift distortions; 
 c) one knows {\it a priori} the scale on which 
substructures are to be expected. We have found that to allow an accurate 
recovery we must have both a significant number of galaxies ($\approx 200$ for 
clusters at  z$\geq 0.4$ or about 800 at 
z$\leq$ 0.4)  
and a limiting magnitude for completeness  $m_B=16$.\\
The only true limitation to our method seems to be the 
necessity of knowing {\it a priori} the scale on which the 
substructure is to be found. This is an intrinsic drawback of the 
method and no improvement in numerical codes based on this technique 
could make up for it.

\end{abstract}


\section{Introduction}
A large fraction of Clusters of galaxies shows a very complex structure,
either in the spatial and velocity distribution of the galaxies themselves
(e.g. West \& Bothun 1990),
or in the X-ray maps (e.g. Sarazin 1988), or in both. 
Although the substructures are often
clearly visible in the optical (e.g. Mellier et al. \shortcite{Mell}) and 
on the X-ray maps (Grebenev et al. 1995) it is not always easy to
make a quantitative analysis of the amount of substructure
present. During the past years, many methods borrowed from
statistical theory have been applied to attempt determining the
number and properties of the subgroups. Most of these methods have been 
applied to 2D angular position galaxy maps or 1D line-of-sight velocity 
distributions. In a recent paper (Gambera et al. 1997, hereafter GPAB) we have tried to
make use of the full 3D information available from a complete 
sample of galaxy projected positions and line-of-sight velocities 
 for Abell 1656 (the Coma cluster) in order to recover and determine some
morphological properties of the substructure. In this paper we
will explain and test the method introduced in GPAB on simulated cluster catalogs 
and we will
determine the minimal requirements which have to be satisfied
in order to apply it with confidence.\\
Our method is based on the use of the Discrete Wavelet
Transform (hereafter DWT), a technique originally introduced in turbulence studies 
(see e.g. Farge 1992).
Sometimes the Continuos Wavelet Transform (hereafter CWT) has been applied
to characterize substructure in galaxy clusters (e.g. Escalera \&
Mazure \shortcite{Esc1}, hereafter EM),
but it has been convincingly shown that the two methods are
unrelated to each other (i.e. CWT is not the continuous limit of
DWT), and that CWT is unsuitable to characterize the
structure of clusters \cite{fp,pf}. The DWT admits a  complete,
compact supported orthogonal bases \cite{da}, and for this
reason it represents a suitable tool for the analysis of {\em finite} samples of objects,
like galaxy catalogs.\\
One point of our 3D DWT method deserves some
clarification, i.e. the fact that we are using heterogeneous data, 
namely the projected
position on the sky and the line-of-sight velocity. Although
many authors have used the distribution function of the
projected velocity  in their statistical analysis of substructure,
one can argue that in the highly nonlinear environment of a
cluster both redshift distortions and relaxation effects destroy
the relationship between line-of-sight (l.o.s.) velocity and distance,
thus making apparently meaningless the use of the l.o.s.
velocity in a context where one tries to isolate substructures in
real space. Suppose however that the substructure we are trying
to identify is made of spatially separated clumps whose inner
velocity dispersion is smaller than the average l.o.s. distance
among the clumps themselves. The velocity of a
galaxy can be written as: $\boldmath{v}_{g}={\rm H}\boldmath{r} +
\boldmath{v}_{pec}$, where H is the Hubble constant and
$\boldmath{v}_{pec}$ measures the deviations induced by random
velocities within the cluster due to relaxation effects and
systematic infall motions. The quantity $\bmath{v}_{pec}$ is a
stochastic variable, described by some probability distribution,
but we know from observations (e.g. the ENACS survey, Mazure et al.
\shortcite{ENACS}) and from simulations \cite{nhvk} that its
moments filtered on cluster scales are finite. So if: ${\rm Hr}>>
{v}_{pec}$ the Hubble term dominates the l.o.s. velocity,
and this quantity can then be used to characterize
substructures whose relative distances within the cluster are
larger than ${v}_{pec}/{\rm H}$.\\
In order to test our 3D DWT method and to determine reliable
confidence limits, 
we have simulated a total number of 20 clusters changing the following parameters:
1. the distance of the cluster from the observer; 2. the number of clumps making
up the cluster; 3. the number of galaxies inside the cluster and the individual
clumps; 4. the mean distance
separating different clumps in the same cluster; 5. 
the completeness of our catalogues. 
The analysis has been  performed  using a parallel code that 
allows a rapid structure detection and morphological
analysis in a 3-D set of data points \cite{pb}. We have also analysed the output of
a N-body simulation, where substructure has a more hierarchical distribution.\\ 
The plan of the paper is as follows: in \S 2 we describe the 
method of analysis,
in \S 3 we describe how we engendered our mock catalogues, which are then
analysed in \S 4. In \S 5 we discuss the statistical robustness of our results 
and finally in \S 6 we report our conclusions.

\section{The method of analysis}

\subsection{Method overview}
The 3D DWT  method can be divided into three steps.
In the first one computes the wavelet matrices on all
the scales investigated: we adopt the
"\'a trous" algorithm to perform this task , as described by \cite{leg1} (hereafter L94).
The same matrices are computed for the data to be analyzed 
and on a
random distribution in the same region of space and on the 
same
grid as the real data. 
On these latter matrices we calculate the threshold 
corresponding
to a certain confidence level in the structure detection. \\
The second step is the segmentation analysis. The aim of this 
analysis
is to have all the connected pixels with a wavelet coefficients
greater than the threshold labelled with an integer number 
different
for every single structure. This allows a rapid identification of
connected regions which are needed as input for the third step.\\
The third and last step of the method is the computation of a morphological
parameter for every structure singled out and of a mean 
morphological parameter for each scale.

In the following of this section we describe the serial version of our algorithm.
A detailed description of the parallel implementation has been given 
in more technical papers \cite{pb,p}.

\subsection{The wavelet transform}

Generally speaking, a wavelet transform is the decomposition 
of basis functions obtained by translation and dilation of
a particular function localized in both physical and frequency
space. 
A characteristic features of this kind of analysis is that
it allows a simultaneous study of both positional and  
scaling properties. Although our method is devised to deal with 3-D
data, for the sake of simplicity  we describe here the
1-D version. The generalization to the 3-D case 
is straightforward.

For a one-dimensional function $f(x)$ the wavelet transform is 
a linear operator that can be
written as:
\begin{eqnarray}
w(s,t) & = & \langle f | \psi \rangle \nonumber \\
& = & s^{-1/2} \int^{+\infty}_{-\infty} f(x) \psi^* \left( \frac{x-t}{s} \right)
dx
\label{eq1}
\end{eqnarray}
where $s( > 0) $ is the scale on which the analysis is performed, $t \in \Re$ is 
the spatial translation parameter and $\psi$ is the 
the Grossmann-Morlet (\shortcite{GM},\shortcite{GM2}) 
analyzing wavelet function 
\begin{equation}
\psi_{(s,t)}(x) = s^{-1/2} \psi \left( \frac{x-t}{s} \right)
\end{equation}
that is spatially centered around position $t$ and on a scale $s$.
The wavelet function  $\psi_{(1,0)}(x)$  is called mother 
wavelet. 
It generates the other wavelet function $\psi_{(s,t)}(x), s>1$.
We follow L94 in the choice of the  
mother wavelet in order to use the {\it \`a trous} algorithm in the
following.
\begin{equation}
\psi(x)=\phi(x)-\frac{1}{2} \phi(\frac{x}{2})
\label{eq:atrous}
\end{equation}
where $\phi$ is the cubic centred B-spline 
function defined by:
\begin{equation}
\phi(x) = \frac {|x-2|^3 - 4|x-1|^3 + 6|x|^3 - 4|x+1|^3 +|x+2|^3} 
{12}
\end{equation}
In order to use the {\it \`{a} trous} algorithm we choose a set of scales which 
are powers of two:  $s=2^r$ and the first scale
always corresponds to the size of 1 pixel.
The scale $s$ in this kind of analysis may be considered as the 
resolution.
In other words, if we perform a calculation on a scale $s_0$, we 
expect the
wavelet transform to be sensitive to structures with typical size 
of about $s_0$
and to be able to reveal them.
The first step of the wavelet matrices computation is the evaluation 
of the coefficient $c(0)$. This is defined as:

\begin{equation}
c(0,t) = \langle f(x) | \phi (x-t) \rangle 
\label{eq:c0}
\end{equation}

On the other scales the coefficients $c$ are given by:

\begin{equation}
c(s,t) = {\frac {1}{2^r} } \langle f(x) | \phi ( \frac{x-t}{s} ) \rangle 
\end{equation}

\noindent
Since the function $\phi$ satisfies:

\begin{equation}
\frac{1}{2^{i+1}} \phi(\frac{x}{2^{i+1}}) = \sum^{n=2}_{n=-2} 
h(n) \phi( \frac{x}{2^i} -n)
\end{equation}

\noindent
for $i \ge 0$, we can write:

\begin{equation}
c(s,t) = \sum_{n=-2}^{2} h(n) c(s-1,t+n2^{r-1}) 
\label{eq:c1}
\end{equation}

\noindent
where $h(n)=\frac{1}{16} C^4_{2-n}$, $C^{m}_{n}$ being the binomial 
coefficients.
Using to eqs.\ref{eq1} and 
\ref{eq:atrous} we can write the following expression for the wavelet
coefficients on the various scales:

\begin{equation}
w(s,t) = c(s,t) - c(s-1,t) 
\label{eq:wa}
\end{equation}

The wavelet analysis associates to each pixel a real number, which 
represents the smoothed local density contrast at a given scale.
At the end of this part our result is a set of matrices of wavelet
coefficients; one matrix for each scale investigated. \\
Even if the histogram of the wavelet coefficients may suggest the presence
of substructure, revealed by asymmetries between the positive 
and negative parts
of the probability distributions (see e.g. figs. 2-4 in GPAB), this
kind of information is only visual and not easily quantifiable and
spatially localizable.

\begin{figure}
\begin{center}
\leavevmode
\epsfxsize=6cm \epsfbox{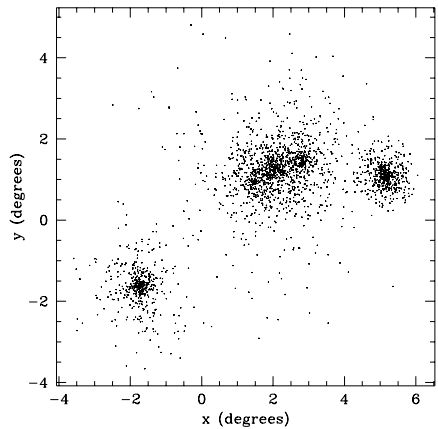}
 \caption{Projected positions of galaxies in a typical mock catalogue
obtained as described in \S 3.1}
\label{fig1}
\end{center}
\end{figure}

\subsection{Thresholding}
The thresholding is made on the wavelet coefficient histogram. 
For an ideal flat background, the wavelet transform coefficients
should be equal to zero.  The existence of structures
at a given scale gives wavelet coefficient with large positive 
values. 
It is however quite obvious that this is strictly true only in an 
ideal
case: a random distribution may have non-zero coefficients 
even if there
are no structures, due to statistical fluctuations. Moreover, the 
statistical
behaviour of the wavelet coefficient is complex due to the 
correlation
among nearby pixels.\\
In order to decide whether a structure detected on a given scale we
need to fix a significance threshold.
We choose it through a classical decision rule. 
We calculate the
wavelet coefficients $w_{ran}(s)$ for each scale of our analysis, 
for a random distribution in the same  region of space of our 
data and on
the same grid.
Then we calculate the probability 
$P[w(s) \le w_{ran}(s)]$  and choose the value $w_{thres}(s)$ so 
that:
\begin{equation}
P[w_{thres}(s) \le w_{ran}(s)]  \le  \epsilon
\end{equation}
Our threshold on the scale $s$ is the value $\nu_{thres} = w_{thres}(s)$. 
For example, a choice for the value of $\epsilon$ of:
\begin{equation}
\epsilon=0.001
\end{equation}
ensures a $99.9 \%$ confidence level in the structure detection.
We have also explored the consequences of an alternative choice for the treshold, i.e.
to fix it in terms of a given number of standard deviations from the variance, but
the final
results are insensitive to this choices.

\subsection{Structure numbering by means of segmentation} \label{sec:seg}
The second step of our analysis is the determination of connected pixels
over a fixed threshold ({\it segmentation}, Rosenfeld
\shortcite{rose}), 
the numbering of the selected structures and
their morphological analysis. 

The segmentation and numbering consists in the exam of the 
wavelet coefficients matrix;  all the pixels 
associated with a wavelet coefficient greater
than the selected threshold are labelled with an integer number. 
All other pixel labels are set equal to zero. 
Then, the same label is associated with all the pixels connected in a single 
structure, in a sequential
way. So, the first structure individuated bears the label '1'  and so on.
We also compute the 
volume and surface of each structure found. 

\subsection{The morphological parameter}
In order to perform a morphological analysis we have to introduce a
morphological parameter that quantifies the sphericity of the structures.
We choose the parameter:
\begin{equation}
L(s)= K(s) \frac{V^2}{S^3}
 \end{equation}
where $V$ is the volume and $S$ is the surface, as in L94, and $K(s)$ is
a parameter that depends on the scale of the analysis. 
We want $L(s)$ to have the following behaviour:  
zero for very filamentary structures and $1$ for spherical ones.
This may be achieved putting $K=36 \pi$, but only for those scales not
affected by the granular nature of the analysis. We choose the value $36 \pi$
only for the scales $s=2^r$ pixels  with $r \ge 2$. 
For the smallest scales the constant $36 \pi $ is not
adequate, since we are close to the grid resolution and the  
geometry of the substructures cannot be spherical. 
Since we want to consider as spherical a one-pixel structure, 
we adopt the values:
\begin{equation}
K(2^r)= \left\{  \begin{array}{ll}  
216 & \mbox{if $r=0,1$} \\  
36 \pi & \mbox{otherwise} 
\end{array} \right.
\end{equation}
Then,  for every detection threshold we calculate the
values:
\begin{equation}
\langle L(s) \rangle = \sum_{i=1}^{N_{obj}} \frac{L(s)}{N_{obj}}
\end{equation}
where $N_{obj}$ is the number of objects detected at scale $s$.

\section{Simulating clusters of galaxies}

\subsection{Mock catalogs}
Our model clusters are engendered by randomly assembling a 
number of clumps,
each characterized by the position of their centers. As 
coordinates we choose two
angular coordinates and the line-of-sight velocity: 
$(\theta,\phi,v_{los})$. Given the
position of the center of each clump 
$(\theta_{cl},\phi_{cl},v_{cl})$, galaxies are
distributed about the center according to a gaussian 
distribution in real space
of width $\sigma_{r}$. Velocities are then specified by adding 
to the Hubble flow
a random component drawn from a maxwellian distribution 
with a given
dispersion $\sigma_{cl}$. Each simulated cluster is then made 
of a collection of
these clumps. This method allows a partial superposition of 
the clumps.\\
To each galaxy in the cluster we also attribute an absolute 
magnitude drawn
randomly from a Schechter luminosity function:
\begin{equation}
\phi(L) = \phi^{\ast}\left( \frac{L}{L_{\ast}}^{-\alpha}\right) e^{-L/L_{\ast}}\frac{dL}{L_{\ast}}
\end{equation}
The values for  $m_{\ast}=-2.5{\rm Log}(L_{\ast}), \phi^{\ast}$ and $\alpha$ are
appropriate to the central region of the Coma cluster \cite{bdglflms}.\\
We show in Fig. 1 the projected positions of galaxies in a typical mock catalogue
obtained with these prescriptions, and in Fig. 2 the corresponding wedge
diagram.
It is evident from these plots that substructure in these mock catalogues is
made of independent groups which overlap each other.

\begin{figure}
\begin{center}
\leavevmode
\epsfxsize=6.cm \epsfbox{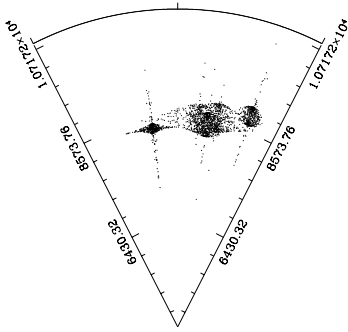}
 \caption{The wedge diagram corresponding to the projected
positions of fig.1}
\label{fig2}
\end{center}
\end{figure}

\subsection{N-body simulation}
We also applied our test to the result of a N-body simulation, where density
distribution is clearly hierarchical. We have considered a $10^{3} h^{-3} Mpc^{3}$
box extracted from a 16-million particles simulation of a 50 $h^{-1} Mpc$ 
box (Antonuccio-Delogu et al., in preparation). Using SKID, a standard 
gravitationally bound groups finder (Stadel et al., in preparation)
we have found 5 groups on a scale of 500 $h^{-1}$ kpc within this box.  

\section{Testing substructure recovery}

\subsection{Is the DWT capable to recover substructure?}
In order to assess the capability of our code to recover substructures we
simulate five clusters of galaxies at a fixed distance made of an increasing 
number of well separated clumps: from only 1 clump to 15. Clusters are
identified by the names C13 (containing 1 clump, 528 galaxies), C14 (3 clumps,
1224 galaxies), 
C1 (5 clumps, 1512 galaxies), C15 (8 clumps, 3456 galaxies) and C16 
(15 clumps, 8750 galaxies). The mean separation between two clumps inside
a cluster is set equal to $2.7 \, h^{-1}$ Mpc.
As one can see from Table \ref{tab1} substructures are well detected. Unfortunately,
an intrinsic drawback of the DWT is the unability to 
distinguish the typical scale of substructures
(Biajoui, private communication). 
Substructures ''created'' by our
simulations are on a typical scale of about $2 \, h^{-1}$ Mpc, 
so we can detect them looking at the right scale.

\subsection{Dependence on the distance from the observer}
We simulate four different clusters of galaxies with distances ranging 
from $8$ to $68 \, h^{-1}$ Mpc with step $20 \, h^{-1}$ Mpc. 
Clusters are identified by the names C1 (set at a distance $68 \, h^{-1}$ Mpc
from the observer), C2 ($48 \, h^{-1}$ Mpc), C3 ($28 \, h^{-1}$ Mpc) and
C4 ($8 \, h^{-1}$ Mpc). All clusters are made of 1512 galaxies divided in
5 clumps with mean separation $2.7 \, h^{-1}$ Mpc.

\begin{table}
\caption{
Number of distinct substructures above the significance threshold detected by our
method. All lengths are expressed in $h^{-1}$ Mpc. The first six columns report,
respectively: Name of the simulated cluster, Magnitude for 100 \% completeness, 
Number of clumps making up the cluster, Distance to the observer, Total
number of galaxies in the cluster, Mean separation among clumps. In the last four
columns the number of substructures detected at scales $0.5 \div 4 h^{-1}$ Mpc.
The clusters marked by an asterisk have been reshuffled $10+10$ times.
}
\label{tab1}
\begin{tabular}{@{}llllllllll} 
N & $m_{B}$ & $N_{c}$  & $d_{o}$  & $N_{g}$  & $d_{i}$  & 0.5 & 1 & 2 & 4 \\
$C1^{\ast}$  & $\infty$  & 5   & 68    &  1512   &  2.7   &  12  & 6   & 5   &  2 \\
$C13$ & $\infty$  &   1 &   68 &  528  & - &   2 & 2   & 2  & 1        \\
$C14$ & $\infty$  &   3 &   68 &   1224 & 2.7   & 5   & 3   & 3   & 2        \\
$C15$ & $\infty$  &   8 &   68 &   3456 &  2.7  &  11  & 7   & 7   & 5     \\
$C16$ & $\infty$  &   15 &  68  &  8750  & 2.7   & 18   & 18  & 14   &  9     
\end{tabular}
\end{table}

As one can see from Table \ref{tab2}, the recovery does not depend on the 
distance from the observer. However at small distances the Hubble flow
and the peculiar velocities of the galaxies becomes comparable, and
we must take into account this effect.
We have repeated the analysis on each out of 20 realizations of the four
clusters obtained by randomly ''reshuffling'' the original 
simulated catalogues with two different method: 1. the first 10 reshufflings
were made by redistributing randomly the redshifts among the galaxies while
keeping the angular coordinates fixed (see Table \ref{resh1}); 
2. the remaining 10 reshufflings were made varying the
redshifts (namely the value of $cz$) of a typical value of 
the galaxies peculiar velocities in a random direction, 
while keeping the angular coordinates fixed (see Table \ref{resh2}).

\begin{table}
\caption{Tests on mock clusters: varying distances. Symbols are as in 
Table 1, but varying the mean distance from the observer.
} 
\label{tab2}
\begin{tabular}{l l l l l l l l l l } 
{$N$} &
{$m_B$} & 
{$N_c$}  &
{$d_o$}  &
{$N_g$}  &
{$d_i$}  &
{$0.5$}   &
{1} &
{2} & 
{4} \\
$C1^*$  & $\infty$  & 5   & 68    &  1512   &  2.7   &  12  & 6   & 5   &  2       \\
$C2^*$ & $\infty$  &   5 &   48 &  1512       &  2.7  &  13  & 7   & 5   & 2       \\
$C3^*$ & $\infty$  &   5 &   28 &  1512     & 2.7    & 13   &  7  &  4  &   2      \\
$C4^*$ &  $\infty$  &   5&    8&  1512     & 2.7   &  13  &  6  & 5   &  2       \\
\end{tabular}
\end{table}

\begin{table}
\caption{Tests on mock clusters: varying richness. Symbols as in Table 1.} 
\label{tab3}
\begin{tabular}{l l l l l l l l l l } 
{$N$} &
{$m_B$} & 
{$N_c$}  &
{$d_o$}  &
{$N_g$}  &
{$d_i$}  &
{$0.5$}   &
{1} &
{2} & 
{4} \\
$C1^*$  & $\infty$  & 5   & 68    &  1512   &  2.7   &  12  & 6   & 5   &  2       \\
$C5$ &$\infty$  &    5 &   68 &  118  &  2.7  &  21  & 2   & 2   &   1       \\
$C6$ &$\infty$  &    5 &   68 &  330  &  2.7  &  18  & 7   & 5   &   2       \\
$C7$ &  $\infty$  & 5 &   68 & 552    &  2.7  &  18  &  6  &  6  & 3         \\
$C8$ &  $\infty$  &   5 &   68 & 7032   &  2.7  &  11  &  5  &  5  &   2       \\
$C9$ &  $\infty$  &   5 &   68 &  14712  & 2.7   & 12   & 5   & 5   &  2       \\
\end{tabular}
\end{table}

In the first case, the average values of the number of structures found
is always smaller than the one found in the original simulated clusters.
This test, already performed 
in GPAB on a catalogue of clusters from the Coma cluster with the same result, strengthens 
our confidence on the physical significance of the
structures detected in Coma.
In the second case the number of structures is nearly the same as in the
original catalogues. We ''followed'' some galaxies labelling them in the
reshufflings and we have seen that the galaxies labelled are always inside
the same substructure for the clusters C1, C2, C3 and reshuffled samples,
but may change substructure for the case of the cluster C4. This happens
because in this latter case the peculiar velocities and the Hubble flow 
become comparable. However it is worth noticing that, although the
galaxies may ''leap'' from one substructure to another, the number of
these is always nearly the same.

\subsection{Dependence on the number of galaxies}
We simulate six clusters of galaxies made of an increasing number
of data points (galaxies) ranging from a minimum of 118 to a maximum
of 14712. 
Clusters are identified by the names C5 (made of 118 galaxies), C6 (330 galaxies), 
C7 (528 galaxies), C1 (1512 galaxies), C8 (7032 galaxies) and C9 (14712 galaxies).
All these clusters are composed of 5 clumps with mean separation $2..7 \, h^{-1}$ Mpc.
It is clear from our results that a minimum number of at
least  200 data points is required to ensure a correct substructure
recovery and that an increasing number of points improves the 
confidence of the analysis. 

\begin{table}
\caption[h]{Tests on mock clusters: varying clump distances. Symbols as in Table 1.
}
\label{tab4} 
\begin{tabular}{l l l l l l l l l l } 
{$N$} &
{$m_B$} & 
{$N_c$}  &
{$d_o$}  &
{$N_g$}  &
{$d_i$}  &
{$0.5$}   &
{1} &
{2} & 
{4} \\
$C1^*$  & $\infty$  & 5   & 68    &  1512   &  2.7   &  12  & 6   & 5   &  2       \\
$C10$ &  $\infty$  &   5 &   68 &1416    &  1.5   & 8   &  4  & 3   & 3       \\
$C11$ & $\infty$  &  5  &   68 &  1512     &  3.7    & 13   & 10   & 7   &5     \\
$C12$ & $\infty$  &   5 &   68 &   1512    & 4.7    & 13   & 9   &  5  &  4      \\
\end{tabular}
\end{table}

\begin{table}
\caption[h]{Tests on mock clusters: varying completeness. Symbols as in Table 1.
} 
\label{tab5}
\begin{tabular}{l l l l l l l l l l } 
{$N$} &
{$m_B$} & 
{$N_c$}  &
{$d_o$}  &
{$N_g$}  &
{$d_i$}  &
{$0.5$}   &
{1} &
{2} & 
{4} \\
$C17$ & 22  &   5 &  68  &  2399  & 2.7   & 8   & 7   & 5   &  2     \\
$C18$ & 20  &   5 &  68  &  2399  & 2.7   & 8   & 7   & 5   &  2     \\
$C19$ & 18  &   5 &  68  &  2340  & 2.7   & 7   & 6   & 5   &  2     \\
$C20$ & 16  &   5 &  68  &  2280  & 2.7   & 5   & 5   & 4   &  1     \\
\end{tabular}
\end{table} 

\subsection{Dependence on the interclumps separation}
Clusters are identified by the names C10 
(mean separation $1.5 \, h^{-1}$ Mpc), C1 ($2.7 \, h^{-1}$ Mpc), 
C11 ($3.7 \, h^{-1}$ Mpc) and C12 ($4.7 \, h^{-1}$ Mpc).
We note no significant variations in the number of substructures detected
as the mean interclumps separation vary, as one can see from Table \ref{tab4}, if the
mean separation is greater than $1.5 \, h^{-1}$ Mpc.

\subsection{Dependence on completeness}
Clusters are identified by the names C17 (100 \% completeness at $m_B=22$),
C18 ($m_B=20$), C19 ($m_B=18$) and C20 ($m_B=16$).
We note a progressive decrease in the number of structures found, as the
magnitude of completeness decrease. However, it is worth noticing that,
till the value $m_B=16$ included, this decrease is not dramatic and the
number of structures detected is very close to the effective number of 
structures composing the clusters.
We can consider this result as a warning to keep in mind while
examining catalogues with low completeness. In this case the number of
substructures detected could be less than the effective number of
substructures. 
\begin{table}
\caption[h]{Tests on mock catalogues: summary. Columns are as in Table 1.
}
\label{tab6}
\begin{tabular}{l l l l l l l l l l } 
{$N$} &
{$m_B$} & 
{$N_c$}  &
{$d_o$}  &
{$N_g$}  &
{$d_i$}  &
{$0.5$}   &
{1} &
{2} & 
{4} \\
$C1^*$  & $\infty$  & 5   & 68    &  1512   &  2.7   &  12  & 6   & 5   &  2       \\
$C2^*$ & $\infty$  &   5 &   48 &  1512       &  2.7  &  13  & 7   & 5   & 2       \\
$C3^*$ & $\infty$  &   5 &   28 &  1512     & 2.7    & 13   &  7  &  4  &   2      \\
$C4^*$ &  $\infty$  &   5&    8&  1512     & 2.7   &  13  &  6  & 5   &  2       \\
$C5$ &$\infty$  &    5 &   68 &  118  &  2.7  &  21  & 2   & 2   &   1       \\
$C6$ &$\infty$  &    5 &   68 &  330  &  2.7  &  18  & 7   & 5   &   2       \\
$C7$ &  $\infty$  & 5 &   68 & 552    &  2.7  &  18  &  6  &  6  & 3         \\
$C8$ &  $\infty$  &   5 &   68 & 7032   &  2.7  &  11  &  5  &  5  &   2       \\
$C9$ &  $\infty$  &   5 &   68 &  14712  & 2.7   & 12   & 5   & 5   &  2       \\
$C10$ &  $\infty$  &   5 &   68 &1416    &  1.5   & 8   &  4  & 3   & 3       \\
$C11$ & $\infty$  &  5  &   68 &  1512     &  3.7    & 13   & 10   & 7   &5     \\
$C12$ & $\infty$  &   5 &   68 &   1512    & 4.7    & 13   & 9   &  5  &  4      \\
$C13$ & $\infty$  &   1 &   68 &  528  & - &   2 & 2   & 2  & 1        \\
$C14$ & $\infty$  &   3 &   68 &   1224 & 2.7   & 5   & 3   & 3   & 2        \\
$C15$ & $\infty$  &   8 &   68 &   3456 &  2.7  &  11  & 7   & 7   & 5     \\
$C16$ & $\infty$  &   15 &  68  &  8750  & 2.7   & 18   & 18   & 14   &  9     
\\
$C17$ & 22  &   5 &  68  &  2399  & 2.7   & 8   & 7   & 5   &  2     \\
$C18$ & 20  &   5 &  68  &  2399  & 2.7   & 8   & 7   & 5   &  2     \\
$C19$ & 18  &   5 &  68  &  2340  & 2.7   & 7   & 6   & 5   &  2     \\
$C20$ & 16  &   5 &  68  &  2280  & 2.7   & 5   & 5   & 4   &  1     \\
\end{tabular}
\end{table}

\subsection{Test on a N-body simulation}
In order to test the method on a real hierarchical structure which has
many possible substructures at many length-scales, we run a N-body 
cosmological simulation of a cubic box of present size $L=50 h ^{-1}$ Mpc
with a number of particles $n=16,666.216$.

The purpose of the simulation is to form a configuration clustered
on several scales. For this reason, we use constrained initial 
conditions. 

The wavelet analysis is performed on these sets of points for three
scales: 50, 100, 200  $h ^{-1}$ kpc. The confidence level of 
detection is $99.5 \%$ for all the scales and the thresholds have
been computed from a random simulation in the same region with the
same number of points. 

As one can see from the result in Table \ref{nbody} our method
detects quite a few structures on these scales. In particular on
a scale of 500 $h^{-1}$ kpc it detects 5 structures. We have
analysed the same simulation using $SKID$ (Stadel et al., in preparation), a tool used to
detect gravitationally bound objects, and found more than 50 objects
when using the same percolation length scale. This apparently
contradictory result is not strange when one thinks that $SKID$
looks for gravitationally bound objects, while our method inspects
the local density field and looks for relatively isolated objects.
This cannot be the case of the output of purely collisionless
N-body simulation, where gravitationally isolated bound groups are rare.
In fact, each gravitationally bound group on a given scale hosts also
non-gravitationally bound particles, which are often irregularly
distributed across many different groups.

This restriction of our method should not however cause problems for
galaxy catalogues, because galaxies are relatively isolated objects even
when they trace an hierarchically clustered density field.

As a matter of fact, we already applied a serial version of the wavelet
code on a catalogue of galaxies for the Coma clusters and the
method has been able to detect hidden hierarchical substructures
on different scales (see GPAB).

\section{Statistical robustness}
Drawing from a wavelet analysis of a catalogue
of a combined angular-redshift distribution 
any conclusion 
about the real phase- and configuration space structure 
requires that one verifies first that the
catalogue does not suffer from any systematic selection biases or
from other types of systematic effects like those induced by
redshift distortions, as described by Reg\"{o}s \& Geller (1989) and 
Praton \& Schneider (1989). About
the latter we notice that they have little significance for a
distant cluster like Coma (GPAB), in which 
the Hubble flow term is
dominant over the peculiar velocity {\em within} the
substructures. These effects may on the other hand affect the
analysis in closer clusters, but our work shows that in the
presence of a significant statistics (that we quantify with 
about 800 galaxies members) these effects can be neglected
too.
One can reasonably argue that because the structures
we find in simulated clusters of galaxies 
are generally well within the nonlinear virialized region, 
on these scales we are probing a region of the phase space
detached from the Hubble flow, where the linearity between 
redshift and distance is
completely lost. On the other hand one also expects that the 
phase-space distribution within the nonlinear region 
should be enough well-mixed within {\rm each} clump (if there are
any) that the substructures detected correspond to substructures
in velocity space.\\

In order to check this latter hypothesis
we have repeated the wavelet analysis in the distance dependence
study (the distance is the parameter mostly affected from redshift effects)
on each of 10 realizations obtained by randomly ``reshuffling'' 
the original catalogue, i.e. redistributing randomly the
redshifts among the galaxies while keeping the angular
coordinates fixed.

\begin{table}
\caption{Results of substructure recovery in totally 
reshuffled clusters. All lengths are expressed in $h ^{-1}$ Mpc;
In columns 1 and 2: Name of the simulated cluster and distance to the observer,
respectively. Columns 3-4: number of substructures detected at 4 scales in the
range: $0.5 \div 4 h ^{-1}$ Mpc.
} 
\label{resh1}
\begin{tabular}{l l l l l l } 
{$N$} &
{$d_o$}&  
{$0.5$} &
{1} &
{2} &
{4} \\
$C1$& 68& $8.1 \pm 1.1$  & $4.3 \pm 0.7$ & $4.6 \pm 0.8$ & $2.0 \pm 0.0$ \\
$C2$& 48& $9.0 \pm 1.2$  & $5.3 \pm 0.8$ & $3.9 \pm 0.7$ & $2.0 \pm 0.0$ \\
$C3$& 28& $9.1 \pm 1.3$  & $2.9 \pm 0.8$ & $3.9 \pm 0.9$ & $2.0 \pm 0.1$ \\
$C4$&  8& $9.4 \pm 1.1$  & $3.2 \pm 0.7$ & $3.8 \pm 0.9$ & $2.1 \pm 0.1$ \\
\end{tabular}
\end{table}

\begin{table}
\caption{Same as in Table \ref{resh1}}
\label{resh2}
\begin{tabular}{l l l l l l } 
{$N$} &
{$d_o$}&  
{$0.5$} &
{1} &
{2} &
{4} \\
$C1$& 68& $11.2 \pm 1.8$  & $5.6 \pm 1.2$ & $5.2 \pm 1.2$ & $2.0 \pm 0.1$ \\
$C2$& 48& $13.0 \pm 1.6$  & $6.8 \pm 1.0$ & $4.6 \pm 0.8$ & $2.1 \pm 0.1$ \\
$C3$& 28& $12.8 \pm 1.8$  & $5.4 \pm 1.2$ & $5.2 \pm 1.0$ & $2.0 \pm 0.1$  \\
$C4$&  8& $12.4 \pm 1.2$  & $6.0 \pm 1.2$ & $4.2 \pm 0.6$ & $2.0 \pm 0.0$ \\
\end{tabular}
\end{table}

The results are consistent with those found by EM
who performed a similar analysis for 2-D catalogues. The average
values of the number of structures is always smaller than
the one found in the original catalogue, showing that the
catalogue itself is probably contaminated by some uncertainty,
probably connected to the arbitrariness in the choice of the
redshift limits, by some background contaminants, etc. 

Another kind od ``reshuffling'' has been performed on the 
same data varying the value of the redshift by a typical peculiar
velocity in a random direction and following some target
galaxies. These may leap from one substructure to another,
but as the number of galaxies is high the morphology of
substructure is not modified.

These tests strengthen our confidence on the physical
significance of most of the substructures that can
be detected by means of this method, particularly
when dealing with a great number of galaxies members or
with fewer galaxies, but in a more distant cluster.

\begin{table}
\caption{Results of substructure recovery in a N-body test:
Number of substructures detected at scales 50, 100, 200 $h^{-1}$ kpc
with corresponding morphological parameters}
\label{nbody}
\begin{tabular}{l l l} 
\multicolumn{3}{c}{\bf N-BODY TEST} \\
scale &
N.obj.&  
morph. \\
\hline
{50} & 5 & 0.70 \\
{100} & 5 & 0.58 \\
{200} & 1 & 0.12 \\
\end{tabular}
\end{table}

\section{Conclusions}
In the last years contradictory conclusions on various methods to detect
significant substructures in clusters of galaxies have been reported 
by several authors (Fitchett \& Webster \shortcite{f},
West et al. \shortcite{W}, Dressler \& Shectman
\shortcite{D}, Mellier et al. \shortcite{Mell}, Slezak 
et al. \shortcite{Sle}, EM, Escalera \& Mazure \shortcite{Esc2}, L94,
Lega et al. \shortcite{leg2}, GPAB). Among the methods 
that have
been tested and used for this purpose, we believe that the most
powerful is the one based on wavelet transforms and in this 
paper
we have investigated its dependence on various parameters 
that characterize 
a cluster of galaxies, like its distance from the observer or the 
number of
clumps and/or of galaxies that makes it and on some selection 
effects.    
According to our analysis the wavelet transforms method is a 
very powerful 
method to recover substructures inside clusters of galaxies, 
rather independently from the many features that may vary in a cluster.
The only serious limitation is due to the necessity of knowing
{\it a priori} the scale on which 
the substructure is to be found. This is an intrinsic 
drawback of the
method and no improvement in numerical codes based on 
this technique could
make up for it.  An interscale connectivity graph can be helpful to
discrimate among the scales at which a physical object show features
(Bijaoui \shortcite{bi}, and private communication) but this technique is presently 
beyond the purpose of our analysis. 
Besides, a significant number of data points is required to perform an 
accurate analysis. We estimate about 200 galaxies to be a good minimum number
to allow a rather accurate recovery in a distant cluster. As the distance decreases,
we need a larger number of galaxy members so that a statistical 
redistribution can compensate for
redshift distortion. For the closest clusters we have taken
into account ($8 h^{-1}$ Mpc) 800 members should be enough.
Obviously, the larger the number of data
points, the more accurate the analysis.
Finally, it is clear that this method can not give any kind of dynamical
information on the clusters investigated and that a companion method, like
high resolution N-body simulations (e.g. Becciani et
al. \shortcite{b}, \shortcite{bec}), 
is required for a more complete and detailed study about 
evolutionary states.

\end{document}

\end{document}